\documentclass[11pt]{article}\usepackage[]{graphicx}\usepackage[]{color}
\makeatletter
\def\maxwidth{ %
  \ifdim\Gin@nat@width>\linewidth
    \linewidth
  \else
    \Gin@nat@width
  \fi
}
\makeatother

\definecolor{fgcolor}{rgb}{0.345, 0.345, 0.345}
\newcommand{\hlnum}[1]{\textcolor[rgb]{0.686,0.059,0.569}{#1}}%
\newcommand{\hlstr}[1]{\textcolor[rgb]{0.192,0.494,0.8}{#1}}%
\newcommand{\hlopt}[1]{\textcolor[rgb]{0,0,0}{#1}}%
\newcommand{\hlstd}[1]{\textcolor[rgb]{0.345,0.345,0.345}{#1}}%
\newcommand{\hlkwb}[1]{\textcolor[rgb]{0.69,0.353,0.396}{#1}}%
\newcommand{\hlkwc}[1]{\textcolor[rgb]{0.333,0.667,0.333}{#1}}%
\newcommand{\hlkwd}[1]{\textcolor[rgb]{0.737,0.353,0.396}{\textbf{#1}}}%

\usepackage{framed}
\makeatletter
\newenvironment{kframe}{%
 \def\at@end@of@kframe{}%
 \ifinner\ifhmode%
  \def\at@end@of@kframe{\end{minipage}}%
  \begin{minipage}{\columnwidth}%
 \fi\fi%
 \def\FrameCommand##1{\hskip\@totalleftmargin \hskip-\fboxsep
 \colorbox{shadecolor}{##1}\hskip-\fboxsep
     \hskip-\linewidth \hskip-\@totalleftmargin \hskip\columnwidth}%
 \MakeFramed {\advance\hsize-\width
   \@totalleftmargin\z@ \linewidth\hsize
   \@setminipage}}%
 {\par\unskip\endMakeFramed%
 \at@end@of@kframe}
\makeatother

\definecolor{shadecolor}{rgb}{.97, .97, .97}
\definecolor{messagecolor}{rgb}{0, 0, 0}
\definecolor{warningcolor}{rgb}{1, 0, 1}
\definecolor{errorcolor}{rgb}{1, 0, 0}
\newenvironment{knitrout}{}{} 

\usepackage{alltt}
\usepackage{amsmath}
\usepackage[english]{babel}
\usepackage{bm}
\usepackage{booktabs}
\usepackage{breakurl}
\usepackage{color}
\usepackage{xcolor,colortbl}
\usepackage{comment}
\usepackage{geometry}
\usepackage{graphicx}
\usepackage{hhline}
\usepackage{natbib}
\usepackage{url}
\usepackage{rotating}
\usepackage{tabularx}
\usepackage{tikz}
\usetikzlibrary{arrows}
\usetikzlibrary{positioning}
\usepackage{tipa}

\definecolor{LightCyan}{rgb}{0.88,1,1}
\definecolor{Gray}{gray}{0.85}

\IfFileExists{upquote.sty}{\usepackage{upquote}}{}
\begin{document}

\begin{center}
  {\bf Autocorrelated errors in experimental data in the language sciences: \\
       Some solutions offered by Generalized Additive Mixed Models} \\
\ \\
\ \\

R. Harald Baayen$^{a,b}$, Jacolien van Rij$^a$, Cecile de Cat$^c$ and Simon Wood$^d$ \\
\ \\
$^a$Eberhard Karls University, T\"{u}bingen, Germany \\
$^b$The University of Alberta, Edmonton, Canada \\
$^c$University of Leeds, UK \\
$^d$University of Bath, UK.
\ \\
\ \\
\date{\today}
\end{center}

\vspace*{2\baselineskip}

\section{Introduction}
\label{sec:1}

A problem that tends to be ignored in the statistical analysis of experimental
data in the language sciences is that responses often constitute time series,
which raises the problem of autocorrelated errors.  If the errors indeed show
autocorrelational structure, evaluation of the significance of predictors in
the model becomes problematic due to potential anti-conservatism of p-values.

This paper illustrates two tools offered by Generalized Additive Mixed Models
({\sc gamm}s) \citep{Lin:Zhang:1999,Wood:2006,Wood:2011,Wood:2013} for dealing
with autocorrelated errors, as implemented in the current version of the fourth
author's {\sc mgcv} package (1.8.9): the possibility to specify an {\sc ar(1)}
error model for Gaussian models, and the possibility of using factor smooths
for random-effect factors such as subject and item.  These factor smooths are
set up to have the same smoothing parameters, and are penalized to yield the
non-linear equivalent of random intercepts and random slopes in the
classical linear framework.

Three examples illustrate the possibilities offered by {\sc gamm}s.  First, a
standard chronometric task, word naming, is examined, using data originally
reported in \citet{Tabak:2010}.  In this task, and similar tasks such as
lexical decision, a participant is asked to respond to stimuli presented
sequentially.  The resulting sequence of responses constitute a time series in
which the response at time $t$ may not be independent from the response at time
$t-1$.  For some participants, this non-independence may stretch across 20
or more lags in time.   Second, a study investigating the pitch contour
realized on English three-constituent compounds
\citep{Koesling:Kunter:Baayen:Plag:2012} is re-examined.  As pitch changes
relatively slowly and relatively continuously, autocorrelation structure is
strongly present.  A reanalysis that brings the autocorrelation under
statistical control leads to conclusions that differ substantially from those
of the original analysis.  The third case study follows up on a model reported
by \citet{DeCat:Baayen:Klepousniotou:2014,DeCat:Klepousniotou:Baayen:2015}
fitted to the amplitude over time of the brain's electrophysiological response
to visually presented compound words.  We begin with a short general introduction
to {\sc gamm}s.

\section{Generalized additive mixed models}

Generalized additive mixed models extend the generalized linear mixed model
with a large array of tools for modeling nonlinear dependencies between a
response variable and one or more numeric predictors.   For nonlinear
dependencies involving a single predictor, thin plate regression splines are
available.  Thin plate regression splines ({\sc tprs}) model the response by
means of a weighted sum of smooth regular basis functions that are chosen such
that they optimally approximate the response, if that response is indeed a
smooth function.  The basis functions of {\sc tprs} have much better
mathematical properties compared to basis functions that are simple powers of
the predictor (quadratic or higher-order polynomials).  Importantly, the
smoother is penalized for wiggliness, such that when fitting a {\sc gamm}, an
optimal balance is found between undersmoothing and oversmoothing.

When a response depends in a nonlinear way on two or more numeric predictors
that are on the same scale, {\sc tprs} can also be used to fit wiggly
regression surfaces or hypersurfaces, approximated by means of weighted sums of
regular surfaces which are again penalized for wiggliness.  When predictors are
not isometric, tensor product smooths should be used.  Tensor product smooths
({\sc tps}) approximate a wiggly surface or hypersurface using as basis
functions restricted cubic splines, again with penalization for wiggliness.  

Interactions of numerical predictors with a factorial predictor can be
accomodated in two ways.  One option is to fit a different wiggly line or
surface for each level of such a factor.  Alternatively, one may want to take
one of the factor levels as reference level, fit a smooth for the reference
level, and then fit difference curves or difference surfaces for the remaining
factor levels.  These difference curves have an interpretation similar to
treatment contrasts for dummy coding of factors: The difference curve for level
$k$, when added to the curve for the reference level, results in the actual
predicted curve for factor level $k$.

When a factor has many different levels, as is typically the case for
random-effect factors, it may be desirable to require the individual smooths
for the different factor levels to have the same smoothing parameter.  Together
with a heavier penalty for moving away from zero, the resulting `factor
smooths' are the nonlinear equivalent of the combination of random intercepts
and random slopes in the linear mixed model.

In what follows, examples are discussed using {\tt R}, which follows
\citet{Wilkinson:Rogers:1973} for the specification of statistical models.
Extensions to the notation for model formulae made within the context of the
package for linear mixed models \citep[{\sc
lme4},][]{Bates:Maechler:Bolker:Walker:2015} and the {\tt mgcv} package for
generalized additive mixed models \citep{Wood:2006,Wood:2011} are explained
where used first.

\section{Time series in a word naming task}

Although there is awareness in the field of inter-trial dependencies in
chronometric behavioral experiments
\citep{Broadbent:1971,Welford:1980,Sanders:1998,Taylor:Lupker:2001}, efforts to
take such dependencies into account are scarce.
\citet{deVaan:Schreuder:Baayen:2007} and \citet{Baayen:Milin:2010} attempted to
take the autocorrelation out of the residual error by including as a covariate
the response latency elicited at the preceding trial.  This solution, however,
although effective, is not optimal from a model-building perspective, as the
source of the autocorrelation is not properly separated out from the other
factors that co-determine the response latency at the preceding timestep.

\begin{figure}

  \includegraphics[width=\textwidth]{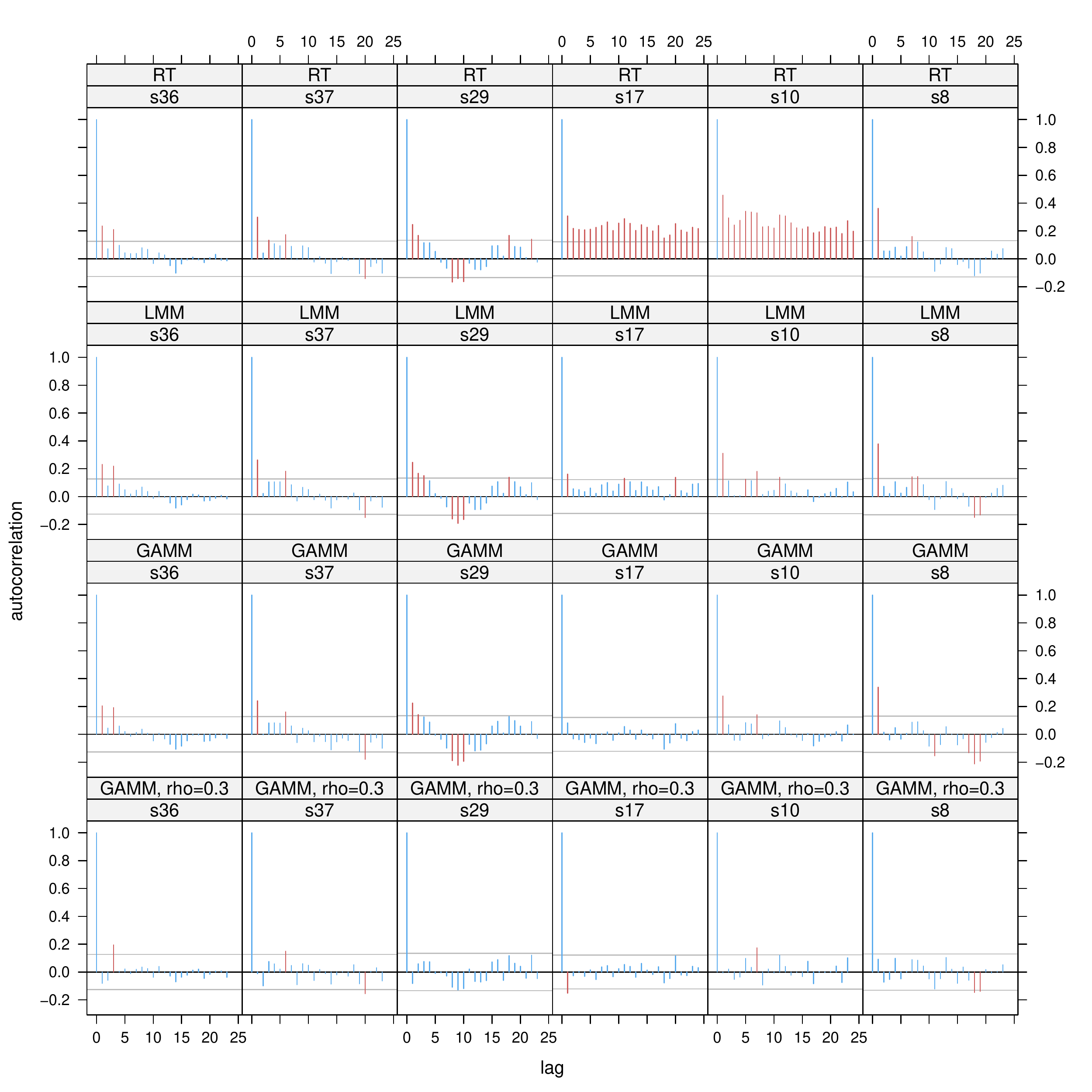}

  \caption{Autocorrelation functions for the residuals of selected participants
  in the word naming task: top: observed response latencies; second row:
  residuals of a linear mixed-effects model with random by-participant
  intercepts and slopes for Trial; third row: residuals of a {\sc gamm} with
  by-participant wiggly curves; fourth row: residuals of a {\sc gamm} with
  by-participant wiggly curves and correction for {\sc ar(1)} with $\rho =
  0.3$. Significant autocorrelations are shown in red, non-significant
  autocorrelations are presented in blue.}

  \label{fig:residsNaming}
\end{figure}

To illustrate the phenomenon, consider data from a word naming study on Dutch
\citep{Tabak:2010}, in which subjects were shown a verb on a computer screen,
and were requested to read out loud the corresponding past (or present) tense
form.  The upper row of panels of Figure~\ref{fig:residsNaming} presents the
autocorrelation function for selected, exemplary, subjects.  The
autocorrelation function presents, for lags 0, 1, 2, 3, \ldots the correlation
coefficient obtained when the vector of responses $\boldmath{v}_1$ at trials 1,
2, 3, \ldots is correlated with the vector of responses $\boldmath{v}_l$ at
trials 1+l, 2+l, 3+l, \ldots ($l >= 0$).  At lag $l=0$, the correlation is
necessarily 1.  As the lag increases, the correlation tends to decrease.  For
some subjects, there is significant autocorrelation at short lags, as
illustrated in the first two panels. The subject in the third panel shows a
``seasonal'' effect, with an initial positive correlation morphing into a
negative correlation around lag 10.  The subjects in the next two panels show a
very different pattern, with autocorrelations persisting across more than 20
lags.

The second row of panels in Figure~\ref{fig:residsNaming} presents the
autocorrelation functions for the residuals of a linear mixed-effects model
fitted to the word naming latencies with random intercepts for item (verb) and
by-subject random intercepts as well as by-subject random slopes for Trial (the
order number of the word in the experimental list, i.e., the variable defining
the time series in this data set).  Using the {\tt lme4} package
\citep{Bates:Maechler:Bolker:Walker:2015} for {\tt R} (version 3.0.2), the
specification of the random effects ({\tt (1 + Trial|Subject)}) requests
by-subject random intercepts, by-subject random slopes for {\tt Trial}, and a
correlation parameter for the random intercepts and slopes.
\begin{knitrout}\small
\definecolor{shadecolor}{rgb}{0.969, 0.969, 0.969}\color{fgcolor}\begin{kframe}
\begin{alltt}
\hlstd{naming.lmer} \hlkwb{=} \hlkwd{lmer}\hlstd{(RT} \hlopt{~}  \hlstd{Regularity} \hlopt{+} \hlstd{Number} \hlopt{+} \hlstd{Voicing} \hlopt{+} \hlstd{InitialNeighbors} \hlopt{+}
                   \hlstd{InflectionalEntropy} \hlopt{+} \hlkwd{poly}\hlstd{(Frequency,} \hlnum{2}\hlstd{)} \hlopt{+} \hlstd{Trial} \hlopt{+}
                   \hlstd{(}\hlnum{1} \hlopt{+} \hlstd{Trial}\hlopt{|}\hlstd{Subject)} \hlopt{+} \hlstd{(}\hlnum{1}\hlopt{|}\hlstd{Verb),}
                   \hlkwc{data} \hlstd{= naming)}
\end{alltt}
\end{kframe}
\end{knitrout}
\noindent
Figure~\ref{fig:residsNaming} (second row) indicates that the thick
autocorrelational structure for subjects 17 and 10 has been eliminated by the
by-subject random regression lines, but the less prominent autocorrelational
structure for the other subjects has remained virtually unchanged.

The third row of panels of Figure~\ref{fig:residsNaming} shows that a {\sc
gamm} with by-subject factor smooths for Trial, replacing the by-subject
straight lines of the linear mixed model yields very similar results.
Using the {\tt bam} function from {\tt mgcv} for {\tt R}, 
the model specification
\begin{knitrout}\small
\definecolor{shadecolor}{rgb}{0.969, 0.969, 0.969}\color{fgcolor}\begin{kframe}
\begin{alltt}
\hlstd{naming.gam} \hlkwb{=} \hlkwd{bam}\hlstd{(RT} \hlopt{~} \hlstd{Regularity} \hlopt{+} \hlstd{Number} \hlopt{+} \hlstd{Voicing} \hlopt{+} \hlstd{InitialNeighbors} \hlopt{+}
                 \hlstd{InflectionalEntropy} \hlopt{+} \hlkwd{s}\hlstd{(Frequency)} \hlopt{+}
                 \hlkwd{s}\hlstd{(Trial, Subject,} \hlkwc{bs}\hlstd{=}\hlstr{"fs"}\hlstd{,}\hlkwc{m}\hlstd{=}\hlnum{1}\hlstd{)} \hlopt{+} \hlkwd{s}\hlstd{(Verb,} \hlkwc{bs}\hlstd{=}\hlstr{"re"}\hlstd{),}
                 \hlkwc{data}\hlstd{=naming)}
\end{alltt}
\end{kframe}
\end{knitrout}

\noindent
requests random intercepts for the verbs ({\tt s(Verb, bs="re")}) and
by-subject wiggly penalized curves for {\tt Trial} ({\tt s(Trial, Subject,
bs="fs", m=1)}, here, {\tt bs="fs"} requests factor smooths with the same
smoothing parameters across subjects, and {\tt m=1} requests shrinkage to
obtain wiggly random effects).

An improvement is obtained by including an autoregressive {\sc ar(1)} process
for the errors:
\begin{equation}
  e_{t} = \rho e_{t-1} + \epsilon_{t},  \hspace*{1em}  \epsilon_{t} \sim {\cal N}(0, \sigma).
\end{equation}
\noindent
This equation specifies that the current error is similar to the preceding
error by a factor $\rho$, with Gaussian noise added. As the current error
depends only on the preceding error, this is a first-order autoregressive
process.  Second-order or higher autoregressive process would also take into
account the error at $t-k, k=2, 3, \ldots$ .  The {\tt bam} function in the {\tt
mgcv} package offers the possibility of taking a first-order autoregressive
process into account by specifying the autoregressive proportionality $\rho$
(with the {\tt rho} directive in the function call) and by supplying a variable
in the data frame, here {\tt NewTimeSeries} (with levels {\sc true, false}),
indicating the beginning of each new time series with the value {\sc true}
(here, the first trial for each subject), to be supplied to the directive {\tt
AR.start} in the call to {\tt bam}:

\begin{knitrout}\small
\definecolor{shadecolor}{rgb}{0.969, 0.969, 0.969}\color{fgcolor}\begin{kframe}
\begin{alltt}
\hlstd{naming.r.gam} \hlkwb{=} \hlkwd{bam}\hlstd{(RT} \hlopt{~} \hlstd{Regularity} \hlopt{+} \hlstd{Number} \hlopt{+} \hlstd{Voicing} \hlopt{+} \hlstd{InitialNeighbors} \hlopt{+}
                     \hlstd{InflectionalEntropy} \hlopt{+} \hlkwd{s}\hlstd{(Frequency)} \hlopt{+}
                     \hlkwd{s}\hlstd{(Trial, Subject,} \hlkwc{bs}\hlstd{=}\hlstr{"fs"}\hlstd{,}\hlkwc{m}\hlstd{=}\hlnum{1}\hlstd{)} \hlopt{+} \hlkwd{s}\hlstd{(Verb,} \hlkwc{bs}\hlstd{=}\hlstr{"re"}\hlstd{),}
                     \hlkwc{rho}\hlstd{=}\hlnum{0.3}\hlstd{,} \hlkwc{AR.start}\hlstd{=naming}\hlopt{$}\hlstd{NewTimeSeries,}
                     \hlkwc{data}\hlstd{=naming)}
\end{alltt}
\end{kframe}
\end{knitrout}
\noindent
There is no automatic procedure for the selection of the value of $\rho$. The
autocorrelation at lag 1 is a good guide for an initial guesstimate, which may
need further adjusting.  When changing $\rho$, it is important not to increase
$\rho$ when this does not lead to a visible reduction in autocorrelation, at
the cost of inflated goodness of fit and warped effects of key predictors.  It
should be kept in mind that an {\sc ar(1)} autocorrelative process is only the
simplest of possible autocorrelative processes that may be going on in the
data, and that hence increasing $\rho$ beyond where it is functional can
distort results.  The final row of Figure~\ref{fig:residsNaming} shows that for
this example, nearly all autocorrelational structure is eliminated with a small
$\rho = 0.3$.  

\begin{figure}
  \centering
  \includegraphics[width=0.8\textwidth]{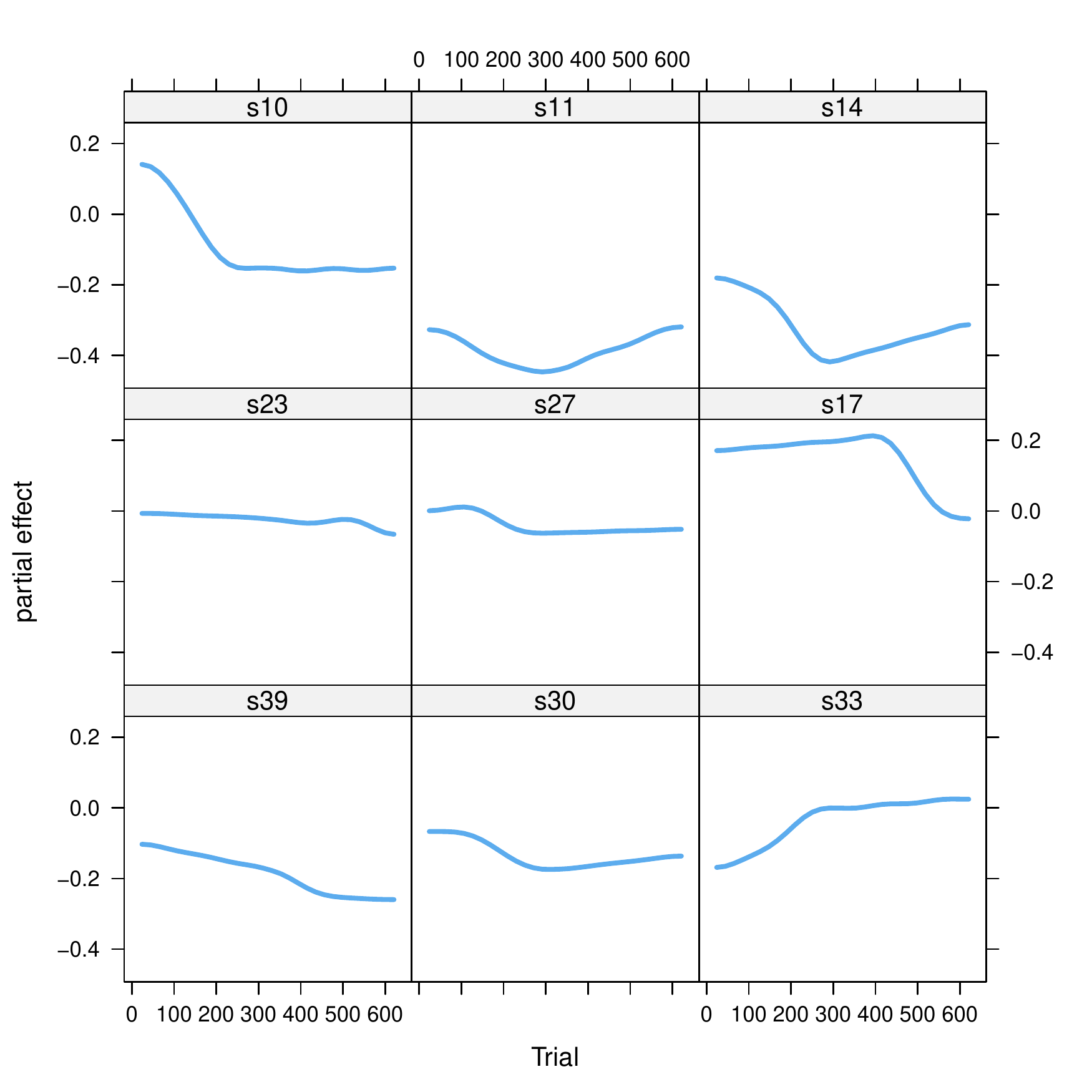}

  \caption{Selected by-subject random wiggly curves for Trial (penalized factor smooths) in the
  {\sc gamm} fitted to word naming latencies.}

  \label{fig:namingWigglies}
\end{figure}

The summary of this model, shown in Table~\ref{tab:gamRT}, shows strong support
for the random effects structure for {\tt Verb} and {\tt Subject}, with large
$t$-values and small $p$-values.\footnote{The parametric coefficients suggest
that regularity is irrelevant as predictor of naming times, that singulars are
named faster than plurals, that words with voiced initial segments have longer
naming times, as do words with a large number of words at Hamming distance 1 at
the initial segment.  Words with a greater Shannon entropy calculated over the
probability distribution of their inflectional variants elicited shorter
response times.  A thin plate regression spline for log-transformed word
frequency suggests a roughly U-shaped effect (not shown) for this predictor.}
Typical examples of by-subject random wiggly curves are shown in
Figure~\ref{fig:namingWigglies}.  These curves capture both changes in
intercept, as well as changes over time.  For some subjects, the changes are
negligible, but for others, they can be substantial, and non-linear.

\begin{table}[ht]
\caption{A {\sc gamm} fitted to log-transformed picture naming latencies ($\rho = 0.3$);
{\tt s}: thin plate regression spline, {\tt fs}: factor smooth, {\tt re}: random effect.} 
\label{tab:gamRT}
\begin{tabular*}{\textwidth}{@{\extracolsep{\fill}}lrrrr}\hline
A. parametric coefficients & Estimate & Std. Error & t-value & p-value \\ 
  Intercept & 6.5531 & 0.0512 & 127.9396 & $<$ 0.0001 \\ 
  Regularity=regular & 0.0093 & 0.0094 & 0.9986 & 0.3180 \\ 
  Number=singular & -0.1147 & 0.0513 & -2.2377 & 0.0253 \\ 
  Voicing=present & 0.0269 & 0.0101 & 2.6734 & 0.0075 \\ 
  Initial Neighborhood Size & 0.0179 & 0.0055 & 3.2499 & 0.0012 \\ 
  Inflectional Entropy & -0.0343 & 0.0159 & -2.1616 & 0.0307 \\ \hline
B. smooth terms & edf & Ref.df & F-value & p-value \\ 
  s(word frequency)   & 4.2914 & 4.6233 & 7.7445 & $<$ 0.0001 \\ 
  fs(Trial, subject)  & 99.4223 & 358.0000 & 5.6670 & $<$ 0.0001 \\ 
  re(verb)            & 190.1753 & 280.0000 & 2.1085 & $<$ 0.0001 \\ 
   \hline
\end{tabular*}
\end{table}

One could consider replacing the factor smooths by by-subject random intercepts,
while at the same time increasing $\rho$.  However, a model such as 
\begin{knitrout}\small
\definecolor{shadecolor}{rgb}{0.969, 0.969, 0.969}\color{fgcolor}\begin{kframe}
\begin{alltt}
\hlkwd{bam}\hlstd{(RT} \hlopt{~} \hlstd{Regularity} \hlopt{+} \hlstd{Number} \hlopt{+} \hlstd{Voicing} \hlopt{+} \hlstd{InitialNeighbors} \hlopt{+}
         \hlstd{InflectionalEntropy} \hlopt{+} \hlkwd{s}\hlstd{(Frequency)} \hlopt{+}
         \hlkwd{s}\hlstd{(Subject,} \hlkwc{bs}\hlstd{=}\hlstr{"re"}\hlstd{)} \hlopt{+} \hlkwd{s}\hlstd{(Verb,} \hlkwc{bs}\hlstd{=}\hlstr{"re"}\hlstd{),}
         \hlkwc{rho}\hlstd{=}\hlnum{0.9}\hlstd{,} \hlkwc{AR.start}\hlstd{=naming}\hlopt{$}\hlstd{NewTimeSeries,}
         \hlkwc{data}\hlstd{=naming)}
\end{alltt}
\end{kframe}
\end{knitrout}
\noindent
provides an inferior fit with an adjusted R-squared of 0.07 (compare 0.36) and
an f{\sc reml} score of 2655 (compare 684).  This suggests that in this data
set, two very different kinds of processes unfold.  One of these processes is
autoregressive in nature, with a relatively small $\rho$.   Possibly, these
autoregressive processes reflect minor fluctuations in attention.  The other
process may reflect higher-order cognitive processes relating to practice and
fatigue, such as exemplified by the fastest subject ({\tt s11}) in
Figure~\ref{fig:namingWigglies}, who initially improved her speed, but then, as
the experiment progressed, was not able to maintain her rapid rate of
responding.

Although these task effects typically are not of interest to an investigator's
central research question, careful modeling of these task effects is important
for the evaluation of one's hypotheses.  For instance, the linear mixed effects
model mentioned previously does not support an effect of inflectional entropy
(Shannon's entropy calculated over the probabilities of a verb's inflectional
variants) with $t = -1.87$, whereas the {\sc gamm} offers more confidence in
this covariate ($t = -2.16$).  However, as we shall see next, predictors may
also lose significance as autocorrelational structure is brought into the
model.

\section{Pitch contours as time series}


\citet{Koesling:Kunter:Baayen:Plag:2012} were interested in the stress patterns
of English three-constituent compounds, and measured the fundamental frequency
of such compounds as realized by a sample of speakers.  In what follows, the
response variable of this study, pitch, is measured in semitones.

\begin{figure}
  \includegraphics[width=\textwidth]{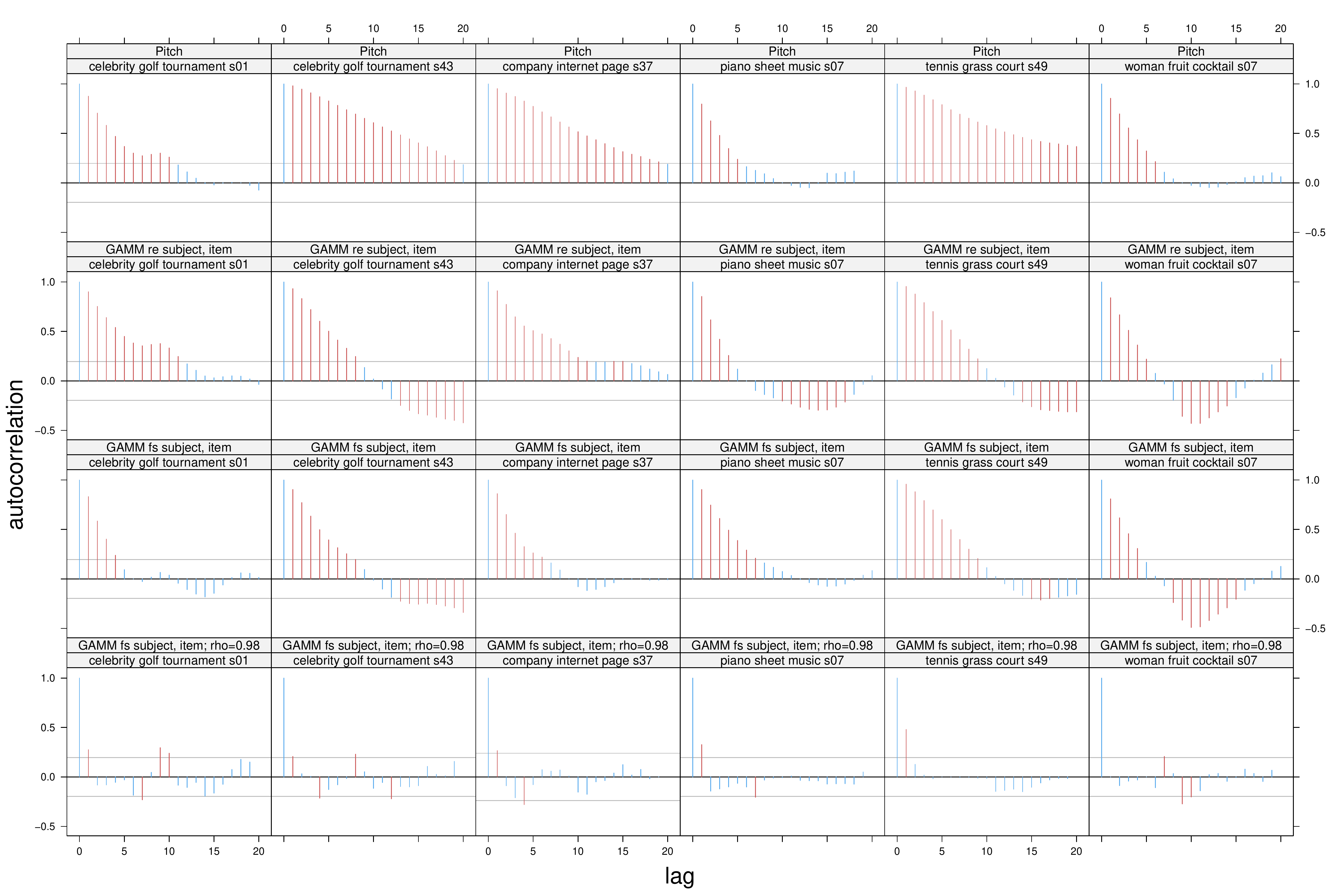}

  \caption{Autocorrelation functions for pitch (in semitones, top row) and
  model residuals (remaining rows) of selected events. Second row: {\sc gamm}
  with by-participant random intercepts and random slopes for {\tt Time} and
  by-compound random intercepts; Third row: {\sc gamm} with by-participant and
  by-compound random wiggly curves; Fourth row: {\sc gamm} with by-compound and
  by-participant random wiggly curves as well as a correction for {\sc ar(1)}
  with $\rho = 0.98$.}

  \label{fig:residsPitch}
\end{figure}

As can be seen by inspecting the top panels of Figure~\ref{fig:residsPitch},
there are autocorrelations in the pitch contours that are much stronger than
those observed for the naming latencies discussed above.   In this figure,
panels represent the autocorrelation functions for selected {\em events}, where
an event is defined as an elementary time series consisting of the pitch
measured at 100 moments in normalized time for the combination of a given
compound and a given speaker.  Whereas for the naming experiment, there are as
many time series as there are subjects, the number of time series in the
present phonetics study is equal to the number of unique combinations of
subjects and compounds ($12 \times 40 = 480$).  


\begin{figure}
  \centering
  \includegraphics[width=\textwidth]{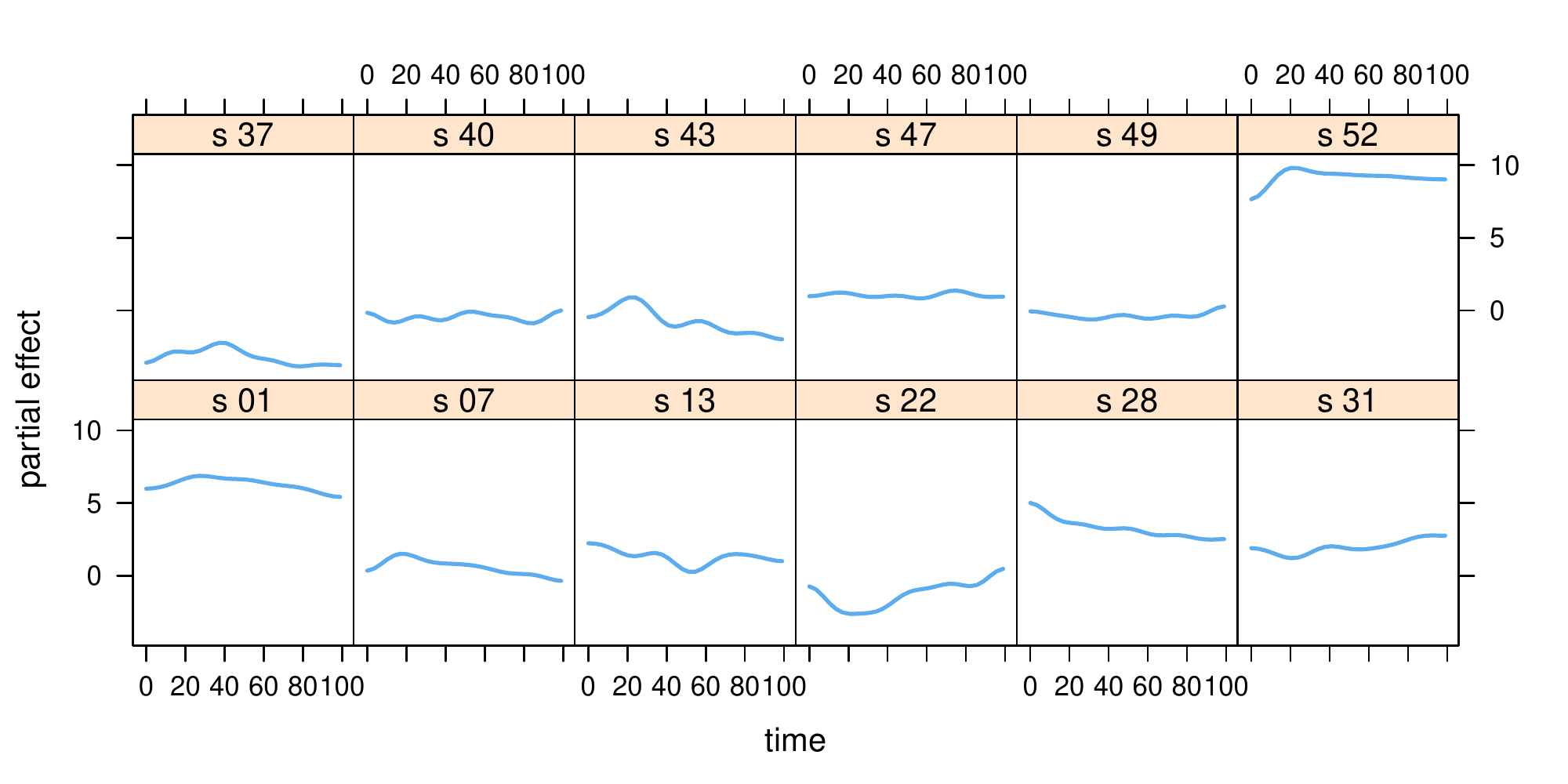}
  \includegraphics[width=\textwidth]{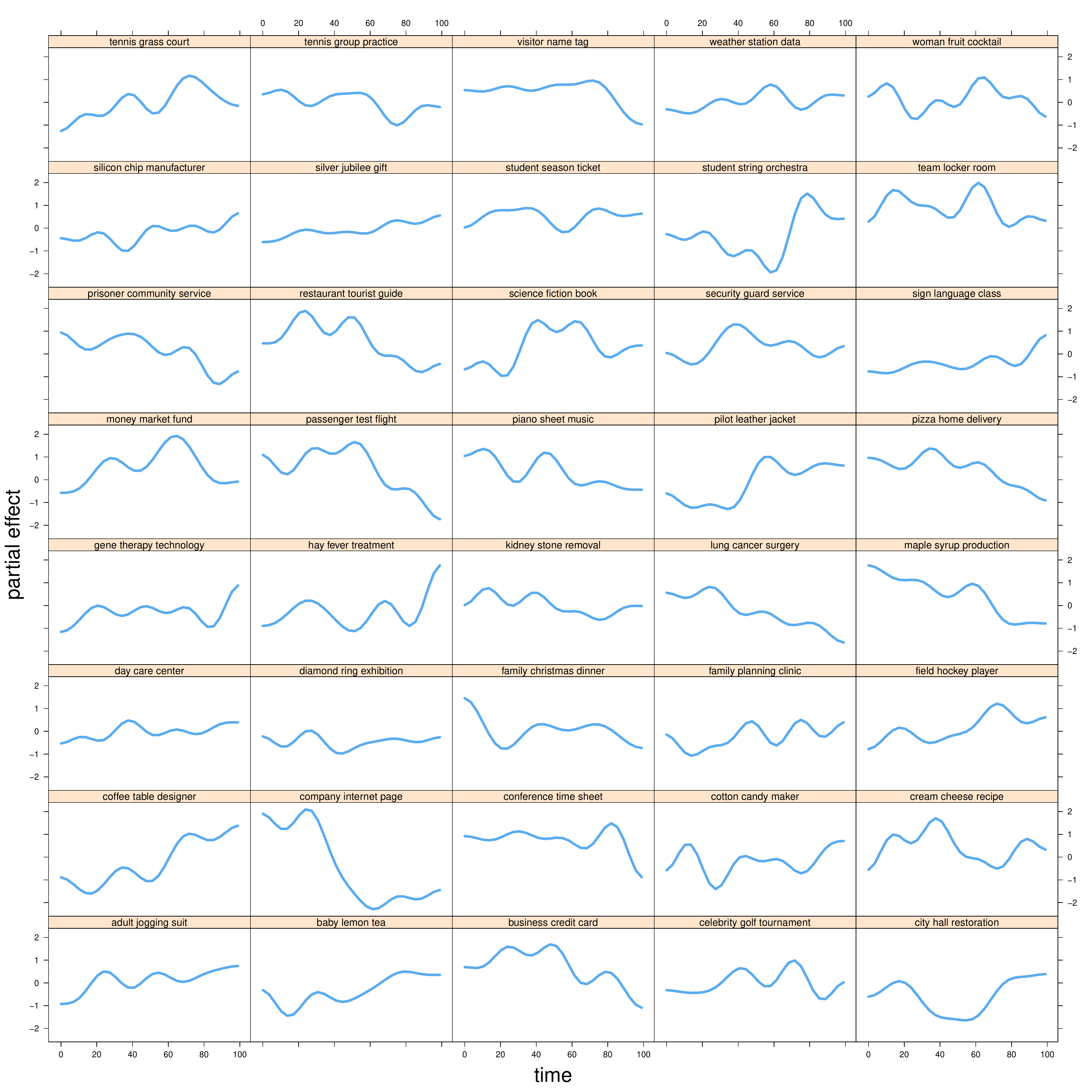}

  \caption{By-speaker (upper trellis) and by-compound (lower trellis) random
  wiggly curves in normalized time in the {\sc gamm} predicting the pitch
  contour for English three-constituent compounds ($\rho =0.98$).}

  \label{fig:pitchWigglies}
\end{figure}

The second row of panels in Figure~\ref{fig:residsPitch} indicates that a model
with by-speaker random intercepts and slopes for (normalized) time does not
succeed in consistently reducing the autoregressive structure of this data.
Some improvement is achieved when by-subject and by-compound random wiggly
curves are added to the model specification (third row of panels), but the
errors are only whitened substantially, albeit not completely, by additionally
including an autoregressive parameter $\rho = 0.98$ (bottom row of
panels).  This fourth model was specified as follows.
\begin{knitrout}\small
\definecolor{shadecolor}{rgb}{0.969, 0.969, 0.969}\color{fgcolor}\begin{kframe}
\begin{alltt}
\hlstd{pitch.gam} \hlkwb{=} \hlkwd{bam}\hlstd{(PitchSemiTone} \hlopt{~} \hlstd{Sex} \hlopt{+} \hlstd{BranchingOrd} \hlopt{+}
  \hlkwd{s}\hlstd{(NormalizedTime)} \hlopt{+} \hlkwd{s}\hlstd{(NormalizedTime,} \hlkwc{by}\hlstd{=BranchingOrd)} \hlopt{+}
  \hlkwd{s}\hlstd{(NormalizedTime, Speaker,} \hlkwc{bs}\hlstd{=}\hlstr{"fs"}\hlstd{,} \hlkwc{m}\hlstd{=}\hlnum{1}\hlstd{)} \hlopt{+}
  \hlkwd{s}\hlstd{(NormalizedTime, Compound,} \hlkwc{bs}\hlstd{=}\hlstr{"fs"}\hlstd{,} \hlkwc{m}\hlstd{=}\hlnum{1}\hlstd{)} \hlopt{+}
  \hlkwd{s}\hlstd{(Compound, Sex,} \hlkwc{bs}\hlstd{=}\hlstr{"re"}\hlstd{),}
  \hlkwc{data}\hlstd{=pitch,}
  \hlkwc{rho}\hlstd{=}\hlnum{0.98}\hlstd{,} \hlkwc{AR.start}\hlstd{=pitch}\hlopt{$}\hlstd{NewTimeSeries)}
\end{alltt}
\end{kframe}
\end{knitrout}
\noindent
{\tt BranchingOrd} is an ordered factor specifying four different compound types
(defined by stress position and branching structure).  The first smooth,
{\tt s(NormalizedTime)}, specifies a wiggly curve for the reference level of this factor.
The second smooth term, {\tt s(NormalizedTime, by = BranchingOrd)},
requests difference curves for the remaining three levels of {\tt BranchingOrd}.\footnote{
For this to work properly, it is necessary to use treatment
contrasts for ordinal factors, in {\tt R}:
{\tt options(contrasts = c("contr.treatment", "contr.treatment"))}.
}
\noindent
A summary of this model is presented in Table~\ref{tab:pitch}.
Figure~\ref{fig:pitchWigglies} clarifies that the variability across speakers
mainly concerns differences in the intercept (height of voice) with variation
over time that is quite mild compared to the variability over time present for
the compounds.

\begin{table}[ht]
\caption{Summary of a {\sc gamm} for pitch as realized on English three-constituent
compounds ($\rho=0.98$); {\tt s}: thin plate regression spline, {\tt ds}:
difference spline, {\tt fs}: factor smooth, {\tt re(compound, sex)}:
by-compound random effects for sex.} 

\label{tab:pitch}
\begin{tabular*}{\textwidth}{@{\extracolsep{\fill}}lrrrr}\hline
A. parametric coefficients & Estimate & Std. Error & t-value & p-value \\ 
  Intercept & 91.3134 & 1.4594 & 62.5689 & $<$ 0.0001 \\ 
  Sex = male & -13.6336 & 1.4649 & -9.3066 & $<$ 0.0001 \\ 
  Branching = LN2 & 0.7739 & 0.4271 & 1.8121 & 0.0700 \\ 
  Branching = RN2 & 0.2415 & 0.3657 & 0.6605 & 0.5089 \\ 
  Branching = RN3 & 0.6460 & 0.4320 & 1.4955 & 0.1348 \\ \hline
B. smooth terms & edf & Ref.df & F-value & p-value \\ 
  s(Time)        & 7.6892 & 7.9403 & 2.7398 & 0.0064 \\ 
  ds(Time, LN2)  & 6.5392 & 7.0804 & 0.6255 & 0.7418 \\ 
  ds(Time, RN2)  & 1.4097 & 1.5555 & 2.4744 & 0.1344 \\ 
  ds(Time, RN3)  & 6.4987 & 7.1541 & 1.9566 & 0.0411 \\ 
  fs(Time, speaker)  & 85.7092 & 105.0000 & 14.2675 & $<$ 0.0001 \\ 
  fs(Time, compound) & 248.5172 & 348.0000 & 3.5294 & $<$ 0.0001 \\ 
  re(compound, sex)  & 19.0558 & 75.0000 & 0.4566 & $<$ 0.0001 \\ \hline
\end{tabular*}
\end{table}

In principle, one could consider fitting a penalized factor smooth to each of
the 480 individual events (time series), although this is currently
computationally prohibitively expensive for the large number of events in the
present study.  The way the model has been specified here is optimistic in the
sense that it assumes that how pitch contours are realized can be factored out
into orthogonal contributions from individual subjects and from individual
compounds.  In a more pessimistic scenario, each event makes its own,
idiosyncratic, contribution to the model's predictions.  In other words, the
present model seeks to capture part of the structure in the elementary time
series by means of crossed wiggly curves `by subject' and `by item'.

Currently, only a single autoregressive parameter $\rho$ can be specified for
all events jointly.  Inspection of the last row of panels of
Figure~\ref{fig:pitchWigglies} suggests that it is desirable to relax the
assumption that $\rho$ is exactly the same for each event.  Although for some
events the autocorrelation function is properly flat already for a moderate
$\rho$, see, e.g., the second panel on the first row ($\rho = 0.4$), events
remain for which autocorrelations persist across several lags.

\begin{figure}
  \centering
  \includegraphics[width=\textwidth]{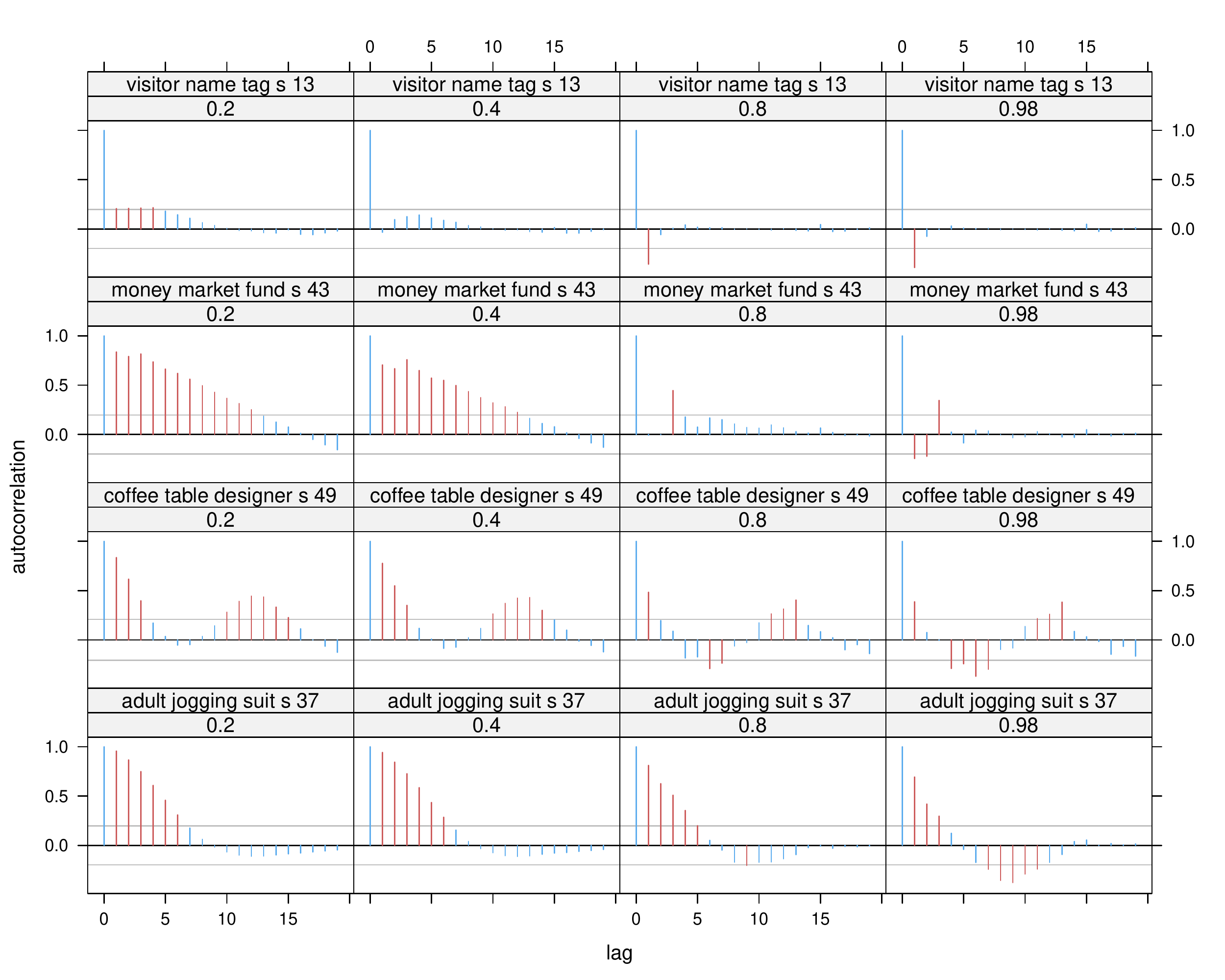}

  \caption{Autocorrelation functions for the residuals of {\sc gamm} models
  with $\rho = 0.2, 0.4, 0.8, 0.98$ (columns) for selected events (rows) where
  the largest value of $\rho$, although for most events optimal, induces
  artifical negative autocorrelations at some lags.}

  \label{fig:rhoProblems}
\end{figure}

Increasing $\rho$ would remove such persistent autocorrelations, but,
unfortunately, at the same time induce artificial autocorrelations for other
events.  This is illustrated in Figure~\ref{fig:rhoProblems}, which presents,
for four events (rows) the autocorrelation function for increasing values of
$\rho$ (columns).   For events with hardly any autocorrelation to begin with
(upper panels), increasing $\rho$ artificially creates a strong negative
autocorrelation at lag 1.  The events in the second and third row show how
increasing $\rho$ can induce artefactual autocorrelations both at shorter lags
(second row) and at longer lags (third row).  The event in the fourth row
illustrates how increasing $\rho$ attenuates but not removes autocorrelation at
shorter lags, while giving rise to new negative autocorrelation at
intermediate lags.  

Although higher-order autoregressive processes might be more appropriate for
many events, they currently resist incorporation into {\sc gamm}s.  Thus, the
analysist is left with two strategies.  The first is to select a value of
$\rho$ that finds a balance between removing strong autocorrelations, while at
the same time avoiding the introduction of artefactual autocorrelation for
events which show little autocorrelation to begin with --- inappropriate use of
$\rho$ may completely obscure the actual patterns in the data.

The second strategy is to remove from the data set those events that show
persistent autocorrelations for the optimal $\rho$ obtained with strategy one.
When refitting the model to the remaining data points yields qualitatively
similar results, it is safe to conclude that the remaining autocorrelational
structure in the original model is not an issue.

\begin{figure}
  \centering
  \includegraphics[width=\textwidth]{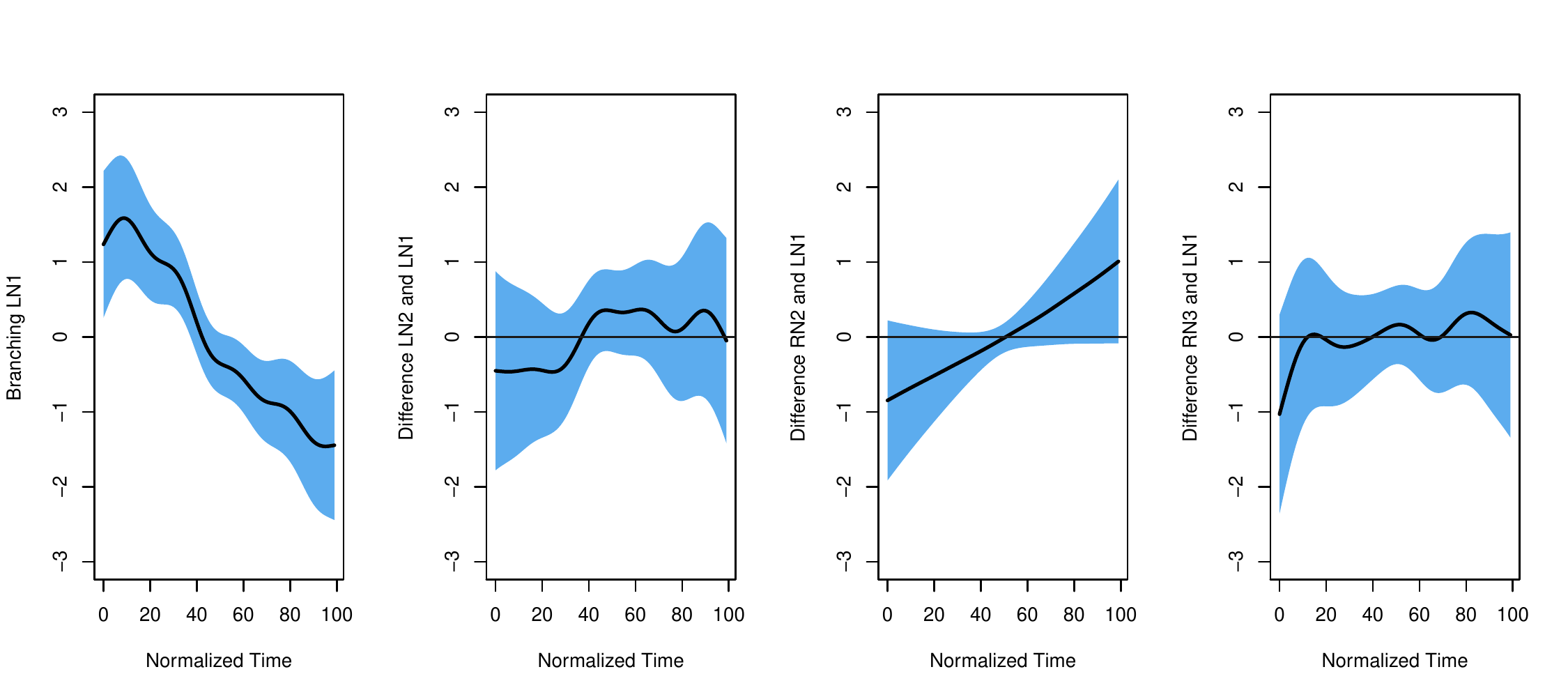}

  \caption{The pitch contour for the {\tt LN1} branching condition, and
  difference curves for the remaining three branching conditions. As
  the confidence regions for the difference curves always contain the
  zero line, there is little support for differences in pitch contour
  as a function of branching condition.}

  \label{fig:pitchCurves}
\end{figure}

Two aspects of the present model are of further interest.  First, the model
includes a thin plate regression smooth for the reference level of compound
type ({\tt LN1}), with difference smooths for the remaining three compound
types.  Inspection of Table~\ref{tab:pitch} reveals only limited support for
significant differences between the pitch contours on the four kinds of
compounds, and inspection of the difference curves (in panels 2--4 in
Figure~\ref{fig:pitchCurves}) clarifies that there is little evidence for
significant differences with the reference curve.  In fact, a simpler model
(not shown) with just a spline for normalized time and no main effect or
interactions involving branching condition fits the data just as well.

The main reason for the absence of the effect of branching condition reported
originally by \citet{Koesling:Kunter:Baayen:Plag:2012} is the inclusion of the
random wiggly curves for compound.  When the factor smooth for compound is
replaced by random intercepts and random slopes for compound, enforcing
linearity, the main effect of branching condition and its interaction with
normalized time is fully significant, just as in the original study.  This
indicates that the variability in the realization of the pitch contours of the
individual compounds is too large to support a main effect of branching condition.

\begin{figure}
  \centering
  \includegraphics[width=\textwidth]{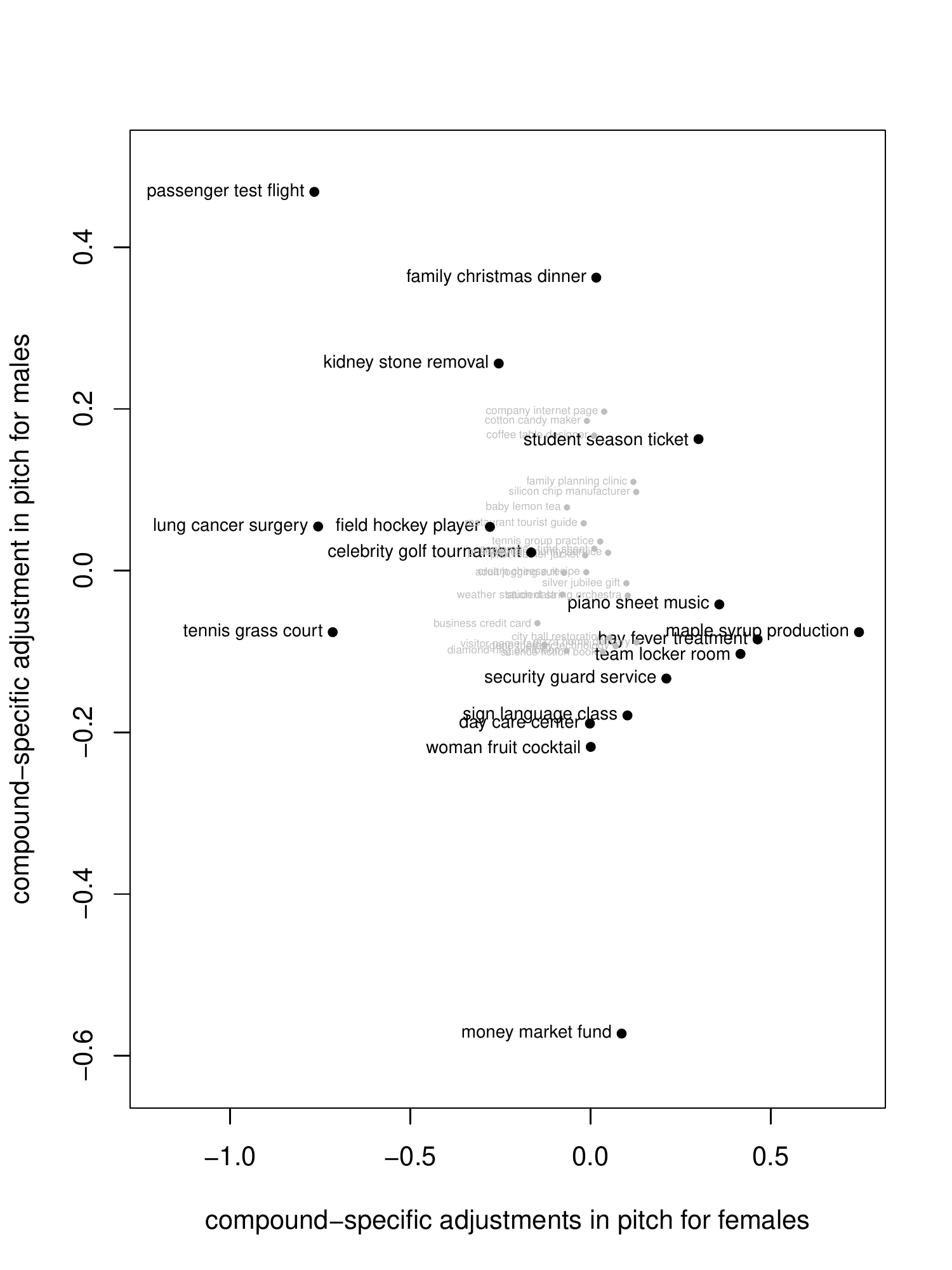}

  \caption{By-compound random contrasts for sex in the {\sc gamm} fitted to the
  pitch contour of English tri-constituent compounds.  Positive adjustments
  indicate a higher pitch.}

  \label{fig:pitchSexesFreq}
\end{figure}

We therefore remove branching condition from the model specification, and
completing the model with a smooth for the frequency of occurrence of the compound, 
\begin{knitrout}\small
\definecolor{shadecolor}{rgb}{0.969, 0.969, 0.969}\color{fgcolor}\begin{kframe}
\begin{alltt}
  \hlkwd{bam}\hlstd{(PitchSemiTone} \hlopt{~} \hlstd{Sex} \hlopt{+} \hlkwd{s}\hlstd{(LogFrequency)} \hlopt{+}
                      \hlkwd{s}\hlstd{(NormalizedTime)} \hlopt{+}
                      \hlkwd{s}\hlstd{(Compound, Sex,} \hlkwc{bs}\hlstd{=}\hlstr{"re"}\hlstd{)} \hlopt{+}
                      \hlkwd{s}\hlstd{(NormalizedTime, Speaker,} \hlkwc{bs}\hlstd{=}\hlstr{"fs"}\hlstd{,} \hlkwc{m}\hlstd{=}\hlnum{1}\hlstd{)} \hlopt{+}
                      \hlkwd{s}\hlstd{(NormalizedTime, Compound,} \hlkwc{bs}\hlstd{=}\hlstr{"fs"}\hlstd{,} \hlkwc{m}\hlstd{=}\hlnum{1}\hlstd{),}
      \hlkwc{data}\hlstd{=pitchc,}
      \hlkwc{rho}\hlstd{=}\hlnum{0.98}\hlstd{,} \hlkwc{AR.start}\hlstd{=pitchc}\hlopt{$}\hlstd{NewTimeSeries)}
\end{alltt}
\end{kframe}
\end{knitrout}
\noindent
we zoom in on the interaction of compound (random-effect factor) by sex
(fixed-effect factor), specified above as {\tt s(Compound, Sex, bs="re")}.
Figure~\ref{fig:pitchSexesFreq} presents a dotplot for the coefficients for the
females on the horizontal axis against the coefficients for the males on the
vertical axis.  Words for which the males tend to raise their pitch are {\em
passenger test flight, family christmas dinner}, and {\em kidney stone
removal}, whereas males lower their pitch for {\em money market fund}.
Females, on the other hand, lower their pitch for {\em tennis grass court, lung
cancer surgery}, and {\em passenger test flight}, but raise their pitch for
{\em maple syrup production, piano sheet music}, and {\em hay fever treatment}.
The two sets of coefficients may even be correlated ($r = -0.31, t(38) =
0.049$), such that where males substantially raise their pitch, females lower
their pitch, and vice versa, possibly reflecting subtle differences in what
topics the different sexes find exciting and unexciting \citep[for pitch
raising as an index of excitement, see,
e.g.,][]{paeschke1999f0,trouvain2000prosody,traunmuller1995frequency}.\footnote{
The details of the coefficients in the present model differ from those obtained in the analysis of
\citet{Baayen:2013}.  Thanks to the factor smooths for subject and compound and
the inclusion of a thin plate regression spline for word frequency, the present
model provides a better fit ({\sc aic} 177077.4 versus 187308), suggesting the
present reanalysis may provide a more accurate window on sex-specific
realizations of compounds' pitch.}

This case study illustrates three methodological points.  First, including
random effect curves (by means of factor smooths) for subjects and items may
lead to substantially different conclusions about the form of smooth terms in
the fixed-effect part of the model specification.   Just as including random
slopes for a factor $X$ may render the main effect of $X$ non-significant in
the context of a linear mixed-effects model, so inclusion of random wiggly
curves for a time series $t$ may render an interaction {\tt s(t, by=X)}
non-significant.  Second, the coefficients of random-effect interactions such
as {\tt Compound} by {\tt Sex} may yield novel insights, especially in the
presence of correlational structure.  Third, when residuals reveal
autocorrelational structure, the {\sc ar(1)} parameter $\rho$ should be chosen
high enough to remove substantial autocorrelational structure, but not so high
that new, artificial autocorrelational structure is artefactually forced onto
the data.

\section{Time series in EEG registration}

Similar to the pitch data, {\sc eeg} data comprise many small time series, one
for each event for which a subject's electrophysiological response to a
particular stimulus is recorded.
\citet{DeCat:Baayen:Klepousniotou:2014,DeCat:Klepousniotou:Baayen:2015} used
English compounds as stimuli, presented in their grammatical order ({\em coal
dust}) and in a manipulated, reversed and ungrammatical order ({\em dust coal})
to native speakers of English as well as advanced Spanish and German learners
of English.  The goal of this study was to clarify whether proficiency and
language background would be reflected in different electrophysiological
processing signatures for these compounds.  For the purposes of the present
study, the specification of the random-effects structure and the measures taken
to bring autocorrelational structure in the residuals under control, and the
effects of the choice of $\rho$ on the fixed-effect predictors and covariates
in the model are of particular interest.  In what follows, the analysis is
restricted to the subset of native speakers of English, and to the {\sc eeg}
at channel C1.\footnote{Data points with an absolute amplitude exceeding 15 $\mu V$,
approximately 2.6\% of the data points, were removed to obtain an approximately
Gaussian response variable.}




The model for these data,
\begin{knitrout}\small
\definecolor{shadecolor}{rgb}{0.969, 0.969, 0.969}\color{fgcolor}\begin{kframe}
\begin{alltt}
\hlstd{eeg.gam} \hlkwb{=} \hlkwd{bam}\hlstd{(Amplitude} \hlopt{~}
  \hlkwd{s}\hlstd{(Time,} \hlkwc{k}\hlstd{=}\hlnum{10}\hlstd{)} \hlopt{+} \hlkwd{s}\hlstd{(Time,} \hlkwc{by}\hlstd{=ConstituentOrder,} \hlkwc{k}\hlstd{=}\hlnum{10}\hlstd{)} \hlopt{+}
  \hlkwd{te}\hlstd{(LogFreqC1, LogFreqC2,} \hlkwc{k}\hlstd{=}\hlnum{4}\hlstd{)} \hlopt{+}
  \hlkwd{te}\hlstd{(LogFreqC1, LogFreqC2,} \hlkwc{by}\hlstd{=ConstituentOrder,} \hlkwc{k}\hlstd{=}\hlnum{4}\hlstd{)} \hlopt{+}
  \hlkwd{s}\hlstd{(LogCompFreq,} \hlkwc{k}\hlstd{=}\hlnum{4}\hlstd{)} \hlopt{+} \hlkwd{s}\hlstd{(LogCompFreq,} \hlkwc{by}\hlstd{=ConstituentOrder,} \hlkwc{k}\hlstd{=}\hlnum{4}\hlstd{)} \hlopt{+}
  \hlkwd{s}\hlstd{(Compound,} \hlkwc{bs}\hlstd{=}\hlstr{"re"}\hlstd{)}\hlopt{+}
  \hlkwd{s}\hlstd{(Trial, Subject,} \hlkwc{bs}\hlstd{=}\hlstr{"fs"}\hlstd{,} \hlkwc{m}\hlstd{=}\hlnum{1}\hlstd{)}\hlopt{+}
  \hlkwd{s}\hlstd{(Time, Subject,} \hlkwc{bs}\hlstd{=}\hlstr{"fs"}\hlstd{,} \hlkwc{m}\hlstd{=}\hlnum{1}\hlstd{),}
  \hlkwc{data}\hlstd{=eegC1,} \hlkwc{family}\hlstd{=}\hlstr{"scat"}\hlstd{,}
  \hlkwc{AR.start}\hlstd{=Start,} \hlkwc{rho}\hlstd{=}\hlnum{0.85}\hlstd{)}
\end{alltt}
\end{kframe}
\end{knitrout}
\noindent
comprises a smooth for time for the compounds presented with their constituents
in the normal order (e.g., {\em goldfish}), and a difference curve for the
condition in which constituent order is reversed ({\em fishgold}).  The model
furthermore takes an interaction of the constituent frequencies into account by
means of a tensor product smooth, as well as the corresponding difference
surface for the reversed order condition.  In the light of the very large
number of observations (207,600), we slightly lowered the upper bound of the
number of basis functions in a given dimension to $k = 4$, in order to avoid
fitting overly wiggly surfaces.  A thin plate regression spline is introduced
to account for the effect of compound frequency, again allowing for a
difference between the standard and reversed word order.  Random intercepts for
compound, and two by-subject factor smooths, one for {\tt Time} and one for the
sequence of trials in the experiment ({\tt Trial}, complete the model
description.  The model summary is given by Table~\ref{tab:gam}.

\begin{table}[ht]
  \caption{Generalized additive mixed model fitted to the electrophysiological
  response of the brain at channel C1 to compound stimuli. Rev: reversed
  constituent order in the compound, Norm: normal order. {\tt s}: thin plate
  regression spline, {\tt te}: tensor product smooth, {\tt re}: random
  intercepts, {\tt fs}: factor smooth. ($\rho = 0.85$)} 
\label{tab:gam}
\begin{tabular*}{\textwidth}{@{\extracolsep{\fill}}lrrrr}\hline
A. parametric coefficients & Estimate & Std. Error & t-value & p-value \\ 
  (Intercept) & 0.0552 & 0.4221 & 0.1308 & 0.8960 \\ \hline
B. smooth terms & edf & Ref.df & F-value & p-value \\ 
  s(Time) & 8.5653 & 8.6645 & 14.6953 & $<$ 0.0001 \\ 
  s(Time):Order=reversed & 1.5768 & 1.9624 & 0.9999 & 0.4139 \\ 
  s(CompFreq) & 1.7242 & 1.7703 & 0.7804 & 0.3172 \\ 
  s(CompFreq):Order=reversed & 2.6384 & 2.8746 & 21.1108 & $<$ 0.0001 \\ 
  te(FreqC1,FreqC2) & 6.5652 & 6.6936 & 4.4840 & 0.0032 \\ 
  te(FreqC1,FreqC2):Order=reversed & 9.6440 & 10.5906 & 10.9593 & $<$ 0.0001 \\ 
  re(Compound) & 99.1995 & 112.0000 & 10.1991 & $<$ 0.0001 \\ 
  fs(Trial,Subject) & 49.5668 & 89.0000 & 12.6940 & $<$ 0.0001 \\ 
  fs(Time,Subject) & 67.4796 & 89.0000 & 8.5343 & $<$ 0.0001 \\  \hline
\end{tabular*}
\end{table}

\begin{figure}
  \centering
  \includegraphics[width=0.8\textwidth]{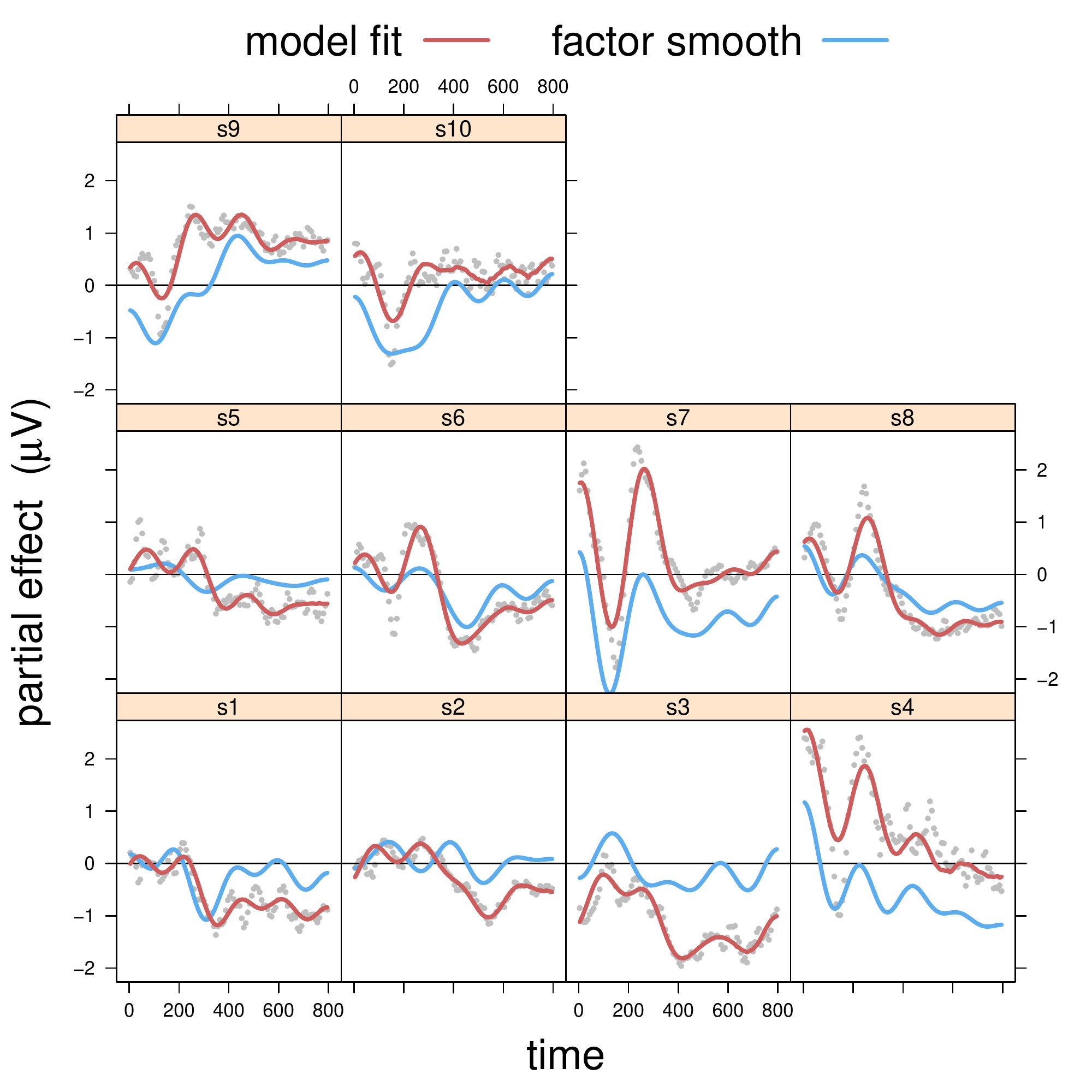}

  \caption{The by-subject factor smooths for {\tt Time} in the {\sc gamm}
  fitted to the {\sc eeg} data.  Dots represent average response times, the
  red lines represent the corresponding average for the model fit, and the 
  blue lines the individual factor smooths.}

  \label{fig:factorSmoothSubject}
\end{figure}

The contributions of the by-subject factor smooths to the model fit is
presented in Figure~\ref{fig:factorSmoothSubject}. The grey dots represent the
by-subject average amplitude for each of the points in time $t = 4, 8, 12,
\ldots$ milliseconds.  The red line shows the average of the model fit for
the same points in time.  The blue lines visualize the by-subject factor
smooths for {\tt Trial}.  Comparing the red and blue lines, it is clear that
a substantial part of the wiggliness of the model fit is contributed by the
factor smooths.  This figure also illustrates the limitations of the factor
smooths: When trends are spiky, as for instance for subjects s5 ad s6 early in
time, a strongly penalized smooth will not be able to fit the data points in
the spike.

\begin{figure}
  \centering
  \includegraphics[width=0.8\textwidth]{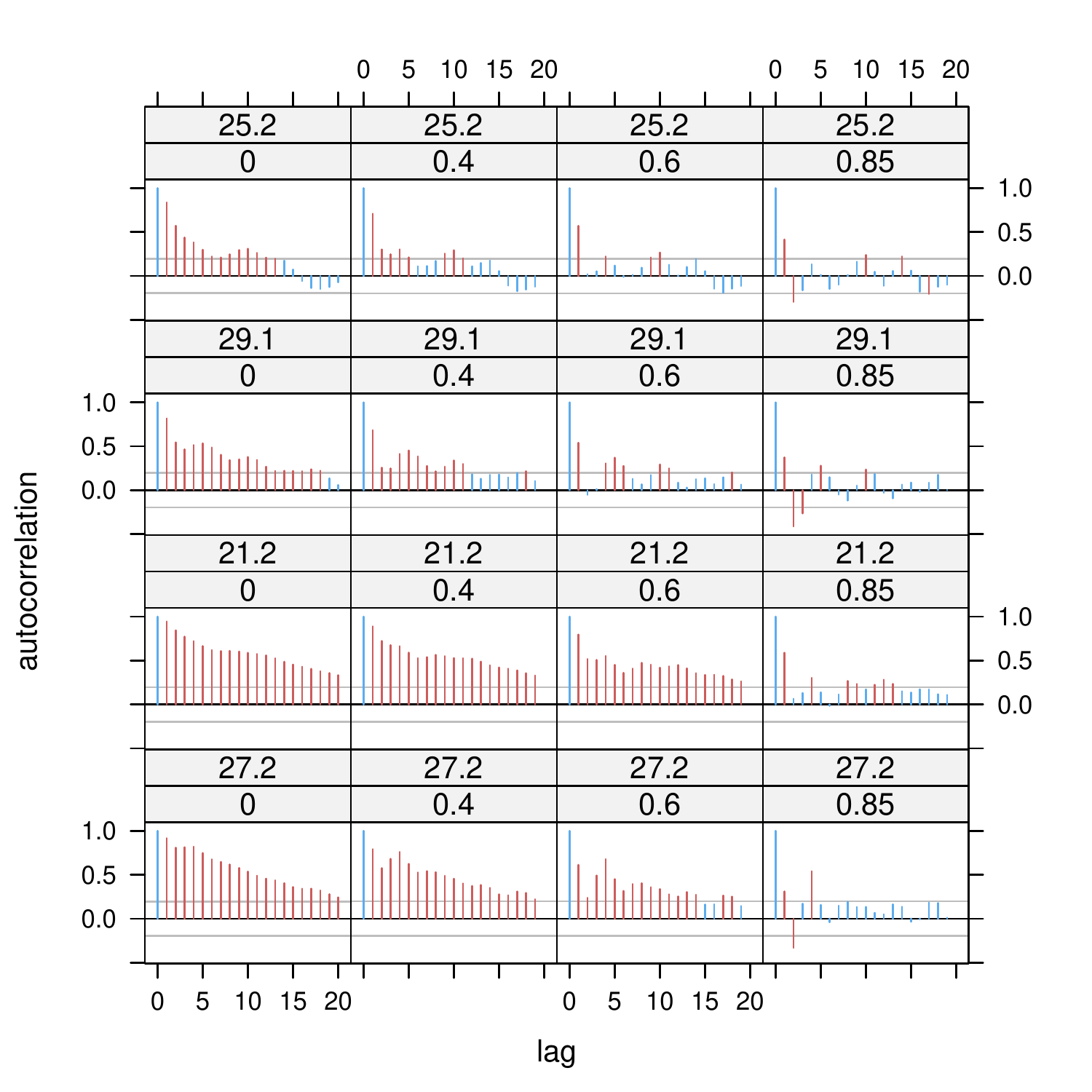}

  \caption{Autocorrelation functions for the residuals of {\sc gamm}s fitted
  to the amplitude of the {\sc eeg} response to visually presented compounds,
  for four events (rows), for $\rho = 0, 0.4, 0.6, 0.85$ (columns).}

  \label{fig:residsEEG}
\end{figure}

\begin{figure}
  \centering
  \includegraphics[width=\textwidth]{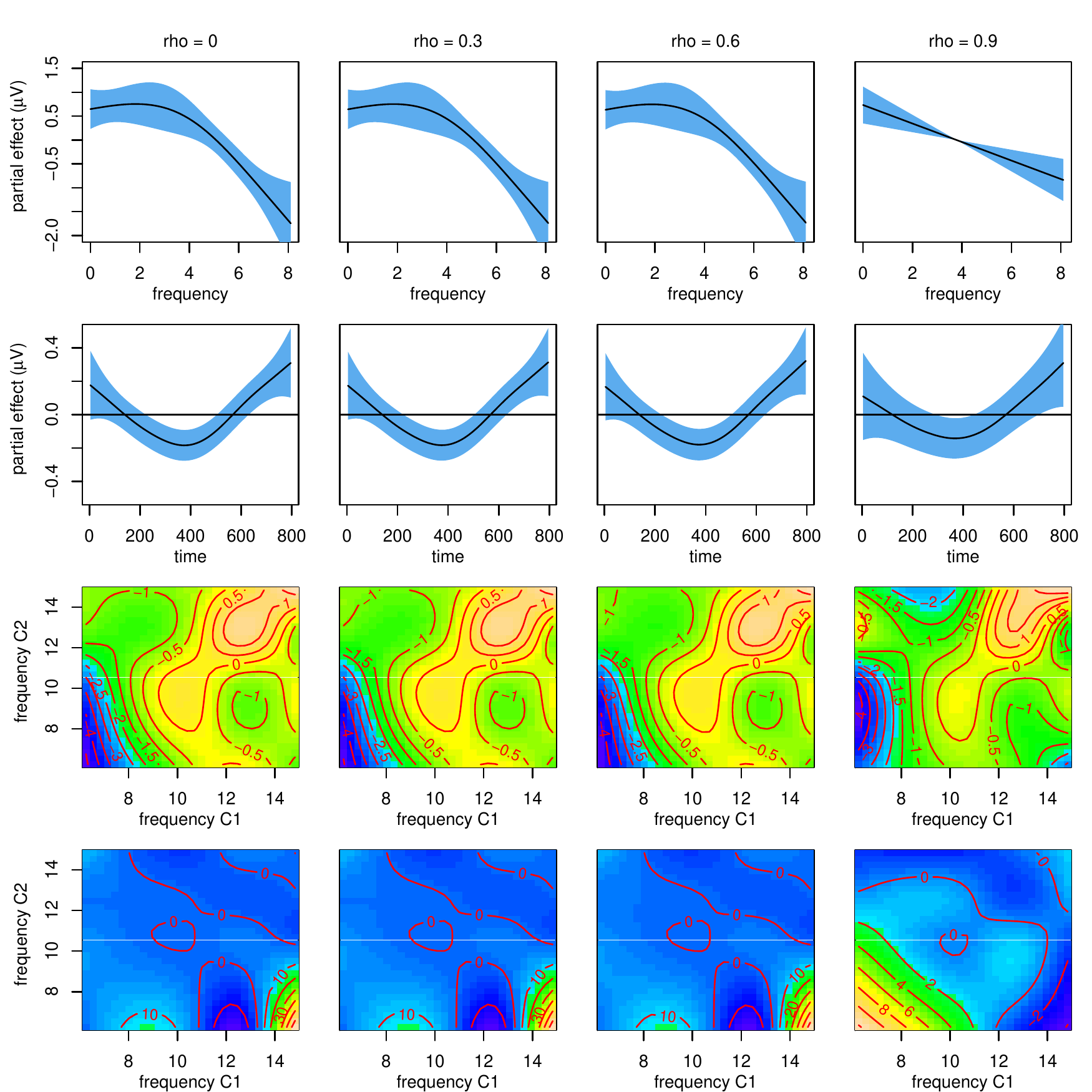}

  \caption{
    The consequences of increasing $\rho$ from 0 to 0.9 (columns) for the
    effect of frequency (top), the difference curve for Time contrasting the
    reversed constituent order with the normal order,  the interaction of the
    frequencies of the first and second constituents (third row), and the
    difference surface for these predictors contrasting the reversed with the
    normal constituent order (fourth row). 
  }

  \label{fig:EEGconsequences}
\end{figure}

Figure~\ref{fig:residsEEG} illustrates, for four events, that $\rho$ cannot be
extended much beyond 0.85 without introducing artefactual negative
autocorrelations.  Interestingly, changing $\rho$ may have consequences for the
predictors of theoretical interest.  Figure~\ref{fig:EEGconsequences}
illustrates this point for four smooths in the model.  The top panels show that
by increasing $\rho$, the effect of word frequency, which at first blush
appears to be nonlinear, becomes a straightforward linear effect.  The second
row of panels clarifies that the difference curve for Time, contrasting the
reversed word-order condition with the normal order, is not trustworthy (see
also Table~\ref{tab:gam}).  The increase in the 95\% confidence interval that
is a consequence of increasing $\rho$ to 0.85, which is required to remove the
thick autocorrelative structure in the residuals (Figure~\ref{fig:residsEEG},
left columns), is noteworthy.

The third and fourth rows of Figure~\ref{fig:EEGconsequences} illustrate that
the regression surface for the frequencies of the compound's constituents
depends on constituent order (a threeway interaction of the frequency of the
first constituent, the frequency of the second constituent, and constituent
order).  The contour plots in the third row show the combined effect of the
constituent frequencies for the normal constituent order, modeled with a tensor
product smooth.  Amplitudes are greater along most of the main diagonal,
suggesting qualitative differences in lexical processing for similar versus
dissimilar constituent frequencies.  For the normal constituent order, this
surface is hardly affected by increasing $\rho$.  This does not hold for the
corresponding difference surface, as can be seen in the bottom row of
Figure~\ref{fig:EEGconsequences}.  In the presence of strong autocorrelations,
autocorrelative noise is incorporated into the tensor surface, leading to
overaccentuated and uninterpretable patterns in the lower right corner of the
partial effect plots. It is only for $\rho=0.9$ that these irregularities
disappear, to give way to a more interpretable difference surface: Amplitudes
in the reversed order condition are reduced compared to the normal constituent
order when both constituents are of a high frequency, whereas amplitudes
increase when both frequencies are low. Thus, this difference surface suggests
that the effect of the constituent frequencies in the normal order is largely
absent when constituent order is reversed.

In summary, removal of autocorrelative structure in the residuals by means of
the $\rho$ parameter for an {\sc ar(1)} error process may have two important
consequences.  First of all, analyses will tend to become more conservative.
Second, the functional form of nonlinear partial effects may change.  In the
present examples, excess wiggliness is removed.

\section{Concluding remarks}

This study illustrates with three examples the potential of generalized
additive mixed models for the analysis of language data: response latencies for
reading aloud, pitch contours of three-constituent compounds, and the
electrophysiological response of the brain to grammatical and ungrammatical
compounds.  

{\sc Gamm}s provide the analyst with two tools for coming to grips with
autocorrelational structure in the model residuals: factor smooths and the {\sc
ar(1)} $\rho$ parameter.  In the standard linear mixed effects model, systematic
changes in how a subject performs over the course of an experiment, or during
an experimental trial with a time-series structure, can only be accounted for
by means of random intercepts and random slopes.   Factor smooths relax this
assumption of linearity, and thereby have the potential to provide much tighter
fits when random-effect factors indeed behave in a non-linear way.  

Autocorrelational structure in the errors may, however, remain even after
inclusion of factor smooths.  For the reaction times revisited in this study,
most of the autocorrelational structure was accounted for by means of factor
smooths for the time series constituted by a participant's responses over the
time course of the experiment.  A mild value of the {\sc ar(1)} correlation
parameter ($\rho = 0.3$) was sufficient to further whiten the residuals.  For
the pitch data, and the same holds for the {\sc eeg} data, inclusion of
by-participant and by-item factor smooths was not successful at all for
removing the autocorrelation.  Here, a high value for the {\sc ar(1)} correlation
parameter was necessary for approximate whitening of the errors. 

Whitening the errors is important for two reasons \citep[see also][for further
discussion]{Baayen:Vasishth:Bates:Kliegl:2015}. First, it protects the analyst
against anti-conservative p-values.  Second, models with whitened errors are
more likely to provide an accurate window on the quantitative structure of the
data.  The analysis of pitch contours provided an example of the inclusion of a
factor smooth rendering a time by fixed-factor interaction non-significant.
Furthermore, whitening {\sc ar(1)} errors may change the functional form of the
effect of predictors of interest.  The analysis of the {\sc eeg} data
illustrated how an effect that initially seemed nonlinear became
straightforwardly linear, as well as a non-linear regression surface that
became simplified and better interpretable thanks to whitening.

%

\bibliography{/home/harald/atransfer/data}
\end{document}